\documentclass[runningheads]{llncs}
\usepackage{graphicx} 
\usepackage{amsmath}
\usepackage{braket}
\usepackage{textcomp}
\usepackage{cleveref}
\usepackage{fancyvrb}
\usepackage{setspace}

\usepackage{booktabs}
\usepackage{multirow}
\usepackage{float}
\usepackage{svg}

\usepackage{graphicx} 
\usepackage{pgfplots}
\usepackage{subcaption} 
\usepackage{amsmath}
\usepackage{tikz}
\usepackage{multirow}
\usepackage{multicol}
\usepackage{threeparttable} 
\usepackage[thmmarks,amsmath]{ntheorem}

\newcommand*\circled[1]{\tikz[baseline=(char.base)]{
            \node[shape=circle,draw, inner sep=0.4pt, line width=0.5pt] (char) {#1};}}

\usepackage{enumitem}
\usepackage{algorithmic}
\usepackage{array}
\usepackage{url}
\usepackage{setspace}

\AtBeginEnvironment{thebibliography}{\footnotesize}

\usepackage{fancyhdr}
\fancypagestyle{firstpage}{
   \fancyhf{} 
   \fancyhead[C]{To appear at the 3rd International Conference on Emergent Quantum Technologies (ICEQT’24) Proceedings} 
}
\begin{document}


\title{NRQNN: The Role of Observable Selection in Noise-Resilient Quantum Neural Networks}


%

\author{Muhammad Kashif and Muhammad Shafique}
%
\authorrunning{M. Kashif et al.}
%
\institute{eBrain Lab, Division of Engineering,\\
Center for Quantum and Topological Systems, NYUAD Research Institute\\
New York University Abu Dhabi, PO Box 129188, Abu Dhabi, UAE 
\email{muhammadkashif@nyu.edu ,  muhammad.shafique@nyu.edu}\\
}

\maketitle
\thispagestyle{firstpage}

\begin{spacing}{0.95}
\begin{abstract}

This paper explores the complexities associated with training Quantum Neural Networks (QNNs) under noisy conditions, a critical consideration for Noisy Intermediate-Scale Quantum (NISQ) devices. We first demonstrate that \emph{Barren Plateaus} (BPs), characterized by exponetially vanishing gradients, emerge more readily in noisy quantum environments than in ideal conditions. 
We then propose that careful selection of qubit measurement observable can make QNNs resilient against noise.
To this end, we explore the effectiveness of various qubit measurement observables, including PauliX, PauliY, PauliZ, and a custom designed Hermitian observable, against three types of quantum noise: Phase Damping, Phase Flip, and Amplitude Damping.
Our findings reveal that QNNs employing Pauli observables are prone to an earlier emergence of BPs, notably in noisy environments, even with a smaller qubit count of four qubits.
Conversely, the custom designed Hermitian measurement observable exhibits significant resilience against all types of quantum noise, facilitating consistent trainability for QNNs up to 10 qubits.
This study highlights the crucial role of observable selection and quantum noise consideration in enhancing QNN training, offering a strategic approach to improve QNN performance in NISQ era.

\end{abstract}
\end{spacing}


%
\vspace{-30pt}
\begin{spacing}{0.85}
\section{Introduction}
The Noisy Intermediate-Scale Quantum (NISQ) devices mark a significant milestone in quantum computing evolution, typically involving 50 to few hundred qubits \cite{Preskill:2018}. Despite their potential to solve classically intractable problems \cite{liang:2023,fan:2023}, NISQ devices are constrained by inherent noise and error rates, which intensify with increasing qubits \cite{Preskill:2018,lau:2022}. This noise complicates quantum coherence and error correction. Nonetheless, NISQ devices are actively being used to investigate quantum applications in cryptography \cite{renner:2023}, financial modeling \cite{herman:2023}, drug discovery \cite{pyrkov:2023} and quantum machine learning (QML) \cite{Biamonte:2017,Benedetti_2019}.


Quantum Neural Networks (QNNs) are a pivotal research area in QML, utilizing Variational Quantum Circuits (VQCs). VQCs are quantum circuits with tunable parameters, which are iteratively optimized during training\cite{cerezo:2021,kashif:ICRC}. This optimization enables QNNs to perform various computational tasks, such as, classification, regression, and pattern recognition \cite{Biamonte:2017,Farhi:2018,kashif:2021}.
However, the scalability of QNNs is significantly challenged by the presence of Barren Plateaus (BPs) in their cost function landscapes \cite{McClean:2018}. BPs are characterized by an exponential decay in the variance of parameter gradients as the number of qubits increases, impeding effective optimization. Addressing this issue is crucial for advancing QNNs and requires focused research efforts.


Various strategies have been proposed to address BPs \cite{Liu_2023,kashif2024resqnets,Kashif:2023alleviating}. The root causes of BPs have been extensively investigated, identifying key factors such as entanglement characteristics \cite{Marrero:2020}, QNN expressibility \cite{kashif_unified}, cost function globality \cite{Cerezo:2021aa,Kashif:2023}, and hardware noise \cite{Wang:2021} as significant contributors.
In the NISQ era, quantum noise is a critical challenge due to its intrinsic nature in current quantum technologies \cite{kashif:2024investigating}. Addressing noise-induced challenges is essential for enhancing QNNs' trainability and realizing the full potential of quantum computing in practical applications \cite{kashif:2024hqnet}.

\vspace{-8pt}
\subsection{Our Contributions}
\vspace{-6pt}
Our main contributions of are summarized below:
\vspace{-10pt}

\begin{itemize} [leftmargin=-.03in]
     \item \textbf{Early Onset of BPs in Noisy Quantum Environments:} We demonstrate the earlier occurrence of BPs in noisy quantum environments. This issue is pivotal as it hinders the learning capabilities of QNNs.

    \item \textbf{Examination of Quantum Noise Types on QNN Performance: } We explore the effects of three major types of quantum noise namely; Phase Damping, Phase Flip, and Amplitude Damping, on QNN's performance. By systematically varying the probabilities of these noise types, we present a nuanced understanding of how different noise scenarios impact QNN's trainability.

    \item \textbf{Comparative Analysis of Qubit Measurement Observables:} We conduct a comprehensive investigation into the impact of various qubit measurement observables, including standard PauliX, PauliY, PauliZ, and a custom-designed Hermitian observable, on the training efficacy of QNNs. This analysis is critical in understanding how these observables effect the overall performance of QNNs under both ideal (noise-free) and noisy conditions.
    
    \item \textbf{Identification of Noise-Resilient QNN Strategies:} A significant contribution of this work is the demonstration that QNNs employing standard Pauli observables are more susceptible to early emergence of BPs, leading to poor training performance in noisy conditions. Conversely, employing custom Hermitian measurement observables shows remarkable resilience against all types of noise, significantly enhancing the robustness and training efficiency of QNNs..

    \item \textbf{Strategic Pathway for Enhancing QNN Training in the NISQ Era:} By showcasing the crucial role of observable selection in QNN training, this paper offers strategic insights for enhancing QNN performance. The identification of a custom Hermitian observable that significantly increases QNN robustness under diverse noise conditions represents a novel approach to designing noise-resilient QNNs.

\end{itemize}



\vspace{-25pt}
\section{Quantum Noise Models} \label{sec:background}
\vspace{-5pt}
Quantum noise, characterized by uncertainty and fluctuations in quantum systems, arises from environmental interference and the challenges of accurately controlling qubits. Unlike deterministic noise in classical systems, quantum noise exhibits probabilistic behavior inherent to quantum mechanics \cite{Yuxuan:2021}. To simulate this in quantum circuits, we incorporate error gates from Pennylane's comprehensive library \cite{Bergholm:2018}, effectively replicating the erratic effects of quantum noise found in NISQ devices. Below, we provide an overview of the different quantum noise types used in this paper.

\vspace{-10pt}
\subsection{Amplitude Damping}
Amplitude damping noise is a type of quantum noise that models energy loss in a quantum system, such as a qubit transitioning from its excited state ($\ket{1}$) to its ground state ($\ket{0}$), with a certain probability. 
%
The mathematical representation of amplitude damping is often realized via two main Kraus operators as discussed below: $K_0$ representing the probability of decay (energy loss) and $K_1$ representing the qubit remaining in its current state without energy loss, where $\gamma \in [0,1]$ represents the amplitude damping probability.
\vspace{-7pt}
$$\footnotesize K_0 = \begin{pmatrix} 1 & 0\\ 0 & \sqrt{1-\gamma}  \end{pmatrix},  K_1 =  \begin{pmatrix} 0 & \sqrt{\gamma}\\ 0 & 0\end{pmatrix}$$


\vspace{-20pt}
\subsection{Phase Damping}
Phase damping, or dephasing, is a type of quantum noise that affects the phase of a qubit's state without causing energy loss. Unlike amplitude damping, which involves transitions between energy levels, phase damping introduces uncertainty in the phase between the computational basis states ($\ket{0}$) and ($\ket{1}$). 
Phase damping noise is modeled using the following Kraus matrices, where $\gamma \in [0,1]$ represents the phase damping probability.

\vspace{-8pt}

$$\footnotesize K_0 = \begin{pmatrix} 1 & 0\\ 0 & \sqrt{1-\gamma}  \end{pmatrix},  K_1 =  \begin{pmatrix} 0 & 0\\ 0 & \sqrt{\gamma}\end{pmatrix}$$ 

\vspace{-20pt}
\subsection{Phase Flip}

Phase flip noise is a type of quantum noise that randomly inverts the phase of a qubit's state while leaving its amplitude unchanged. 
This noise leaves the $\ket{0}$ state unaffected but inverts the phase of the $\ket{1}$ state. Mathematically, phase flip noise is represented by the following matrix, where $p \in [0,1]$denotes the probability of a phase flip:

\vspace{-8pt}
$$\footnotesize K_0 = \sqrt{1-p} \begin{pmatrix} 1 & 0\\ 0 & 1 \end{pmatrix}, K_1 = \sqrt{p} \begin{pmatrix} 1 & 0\\ 0 & -1 \end{pmatrix}$$ 


\vspace{-15pt}

\section{NRQNN Methodology}
\vspace{-8pt}
Our analysis primarily utilizes the hardware-efficient ansatz, a popular approach in NISQ applications characterized by sequences of single and two-qubit gates. The general form of these circuits is given by the following Equation:
\vspace{-8pt}
\begin{equation}\label{eq1}
    U(\theta) = \prod_{i=1}^N U_{\text{ent}}U_{\text{rot}}(\theta_i)
\end{equation}

\begin{figure*}[h]
    \hspace{-30pt}
    \includegraphics[scale=0.25]{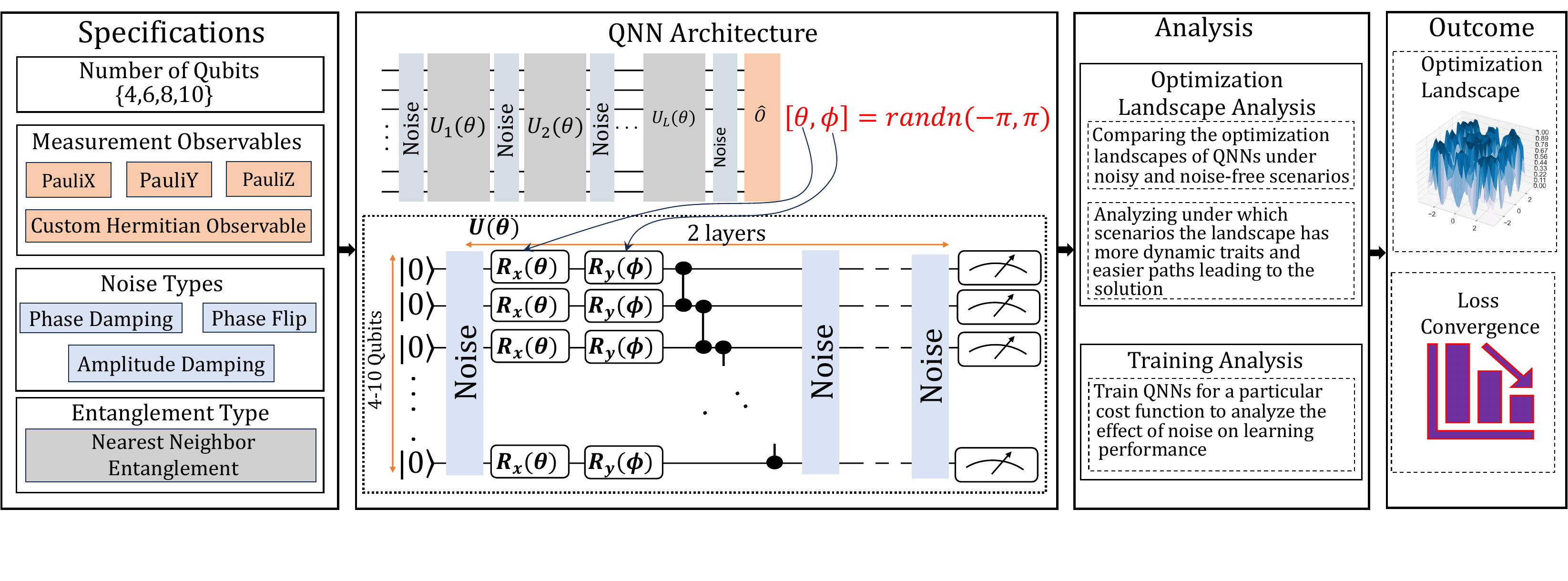}
    \vspace{-35pt}
    \caption{\footnotesize Detailed methodology highlighting key steps for the analysis of noise impact on the trainability of QNNs with different qubit measurement strategies. The quantum circuits used in QNN design are constructed with 4 to 10 qubits. Each qubit has $R_(\theta)$ and $R_y(\phi)$ gates applied on it, and entanglement was achieved between neighboring qubits via the CZ gate. The evaluation metrics used are optimization landscapes and cost function convergence.}
    \label{fig:methodology}
\end{figure*}

\vspace{-4pt}

where $U_{\text{ent}}$ represents a two-qubit gate(s) used for qubit entanglement and $U_{\text{\text{rot}}}(\theta_i)$ represents a single qubit parameterized gate, whose parameters are optimized during training, and $N$ is the total number of repetitions of the quantum circuit until measurement.
A detailed overview of our methodology is depicted in Fig. \ref{fig:methodology}.

\vspace{-10pt}

\subsection{\textbf{Specifications}} \label{sec:specs}

\vspace{-5pt}
\paragraph{\textbf{Number of Qubits.}} 
An important aspect of our study involves examining the trainability of QNNs in relation to the phenomenon of BPs. To comprehensively assess this, we progressively increase the number of qubits, i.e., $Q=\{4,6,8,10\}$. This stepwise increase helps in understanding that how the increase in qubit count influences the optimization landscape, particularly in the context of BPs and their impact on trainability under noise and noise-free environments.

\vspace{-7pt}
\paragraph{\textbf{Types of Observables.}} 
We used different observables for qubit measurement which are PauliZ, PauliX, PauliY and a customized Hermitian measurement observable. 

\vspace{-7pt}

\paragraph*{\textbf{PauliX.}} The PauliX observable, denoted by the symbol $\sigma_x$ or $X$ flips the state of a qubit from $\ket{0}$ to $\ket{1}$ and vice versa, making it fundamental for qubit state manipulation. The eigenvalues of this matrix are $+1$ and $-1$, corresponding to these flipped states.
%
%
\vspace{-5pt}

\paragraph*{\textbf{PauliY.}}
The PauliY observable, denoted by the symbol 
$\sigma_y$ or $Y$, provides information about the phase relationship between the basis states $\ket{0}$ and $\ket{1}$. The PauliY gate flips the state of a qubit from $\ket{0}$ to $i\ket{1}$ and $\ket{1}$ to $-i\ket{0}$. The eigenvalues of this matrix are $+1$ and $-1$, which are related to these phase-adjusted states. 

%
\vspace{-5pt}

\paragraph*{\textbf{PauliZ.}}
The PauliZ measurement observable, often denoted by the symbol $\sigma_z$ or $Z$ differentiates between the two basis states, typically denoted as $\ket{0}$ and $\ket{1}$. The eigenvalues of this matrix are $+1$ and $-1$, corresponding to these basis states. A measurement resulting in +1 indicates the system is in the $\ket{0}$ state, and $-1$ indicates the $\ket{1}$ state.
\vspace{-6pt}
$$\hspace{-8pt} \sigma_x = X = \begin{pmatrix} 0 & 1 \\ 1 & 0 \end{pmatrix}, \sigma_y = Y = \begin{pmatrix} 0 & -i\\ i & 0 \end{pmatrix}, 
 \sigma_z = Z = \begin{pmatrix} 1 & 0 \\ 0 & -1 \end{pmatrix}$$
 
\vspace{-19pt}
\paragraph*{\textbf{Customized Hermitian Observable.}}
The customized Hermitian measurement observable that we have used is a specialized observable, constructed as a Hermitian matrix $H$, tailored to the dimensions of the quantum system under investigation. 
Typically, for a system of $n$ qubits, $H$ is represented by a $2^n \times 2^n$  matrix, initially populated with zeros in all its elements. A distinct modification is then introduced: the diagonal elements of first half ($2^n/2$) rows of $H$ are set to 1, as shown below:
\vspace{-5pt}

\[
\begin{array}{c}
\left[ \begin{array}{ccccccc}
1 & 0 & 0 & 0 & \cdots & 0 & 0 \\
0 & 1 & 0 & 0 & \cdots & 0 & 0 \\
\vdots & \vdots & \vdots & \vdots & \ddots & \vdots & \vdots \\
0 & 0 & 0 & 0 & \cdots & 0 & 0 \\
\vdots & \vdots & \vdots & \vdots & \ddots & \vdots & \vdots \\
0 & 0 & 0 & 0 & \cdots & 0 & 0 \\
\end{array} \right]
\end{array}
\]

Each Pauli matrix (X, Y, Z) has eigenvalues of $+1$ and $-1$, whereas $H$ has $1$ and $0$ as eigenvalues. It highlights a situation where half of the basis states have an expectation value of $1$ and the other half have an expectation value of $0$ The observable $H$ is designed while keeping in mind the following:
\vspace{-8pt}
\begin{itemize}
    \item \textbf{Alignment with Cost Function:} It directly correlates with the cost function that we have used, as described in Eq. \ref{eq:CF}, and betters capture the overlap between the desired state and the current state of the PQC.
    Since it emphasizes the identity part (the first half of the rows in $H$ matrix) of the function while disregarding the state $\ket{00\cdots0}$, it aligns more closely with the objective, providing a clearer signal for the optimization process.

    \item \textbf{Reduced Noise Sensitivity:} Noisy quantum circuits introduce errors that can affect measurements. The observable $H$, by focusing on the probability distribution of a large subset of states (the first half), can average out some of these noise effects, providing a more stable training. In contrast, Pauli observables, which have a more fine-grained structure, might be more sensitive to specific noise patterns.

    \item \textbf{Measurement Statistics:} The cost function derived from $H$ can benefit from better statistical properties. Measuring an observable that effectively averages over a larger subset of the state space can reduce variance in the measurement outcomes, leading to more reliable and stable training.

 
\end{itemize}

\paragraph{\textbf{Noise Types.}} 
\vspace{-10pt}
We train the QNNs under different noisy conditions. To this end, we consider some common quantum noise types named as; phase damping, phase flip and  amplitude damping (details in Section \ref{sec:background}), to analyze their impact on the learning performance of QNNs under consideration. 


\paragraph{\textbf{Entanglement Type.}} 
\vspace{-6pt}
Keeping in mind the limitations of NISQ devices, we used the QNNs with nearest neighbor entanglement, i.e., only the adjacent qubits are entangled.  

\vspace{-6pt}

\subsection{\textbf{QNN Architecture}} \label{sec:method_QNN_arch}
Once the required set of specifications are defined, we then contruct the QNNs. For quantum layers design, we use the hardware-efficient ansatz of the form as shown in Eq. \ref{eq1}. 
Below we present the step-by-step details of our methodology:

\begin{enumerate}
    \item The first step is to define and intialize the qubits. For every qubit number $\in Q$, the qubits are initialized on ground state:
    \vspace{-10pt}
    $$\ket{\psi_0} = \ket{0}^{\otimes n}$$

    \item Once the qubits are initialized, the next is to apply the unitary transformations. We apply two parameterized gates on each qubit:
    \vspace{-8pt}
    $$U_{rot} = \prod_{i=1}^{n} R_y{(\phi_i)} R_x{(\theta_i)}$$

    where $\theta_i$ and $\phi_i$ are the rotation angles for the $i^{th}$ qubit. The (tuneable) rotation parameters are randomly initialized between $(-\pi, \pi)$. 
    Afterwards, the nearest neighbor qubit are entangled:
    \vspace{-10pt} 
    $$ U_{ent} =  \prod_{j=1}^{n-1} CZ_{j, j+1}$$
\vspace{-10pt}

    
    \item Given the above gate specifications, the final QNN is of the form:
    \vspace{-8pt}
     $$ U(\theta) =  U_{ent} U_{rot}$$
     
    \vspace{-20pt}
    
    \begin{equation}\label{eq:train_PQC}
    U(\theta) = \prod_{l=1}^{L}\left(\prod_{j=1}^{n-1}\left(CZ_{j,j+1}\right).\prod_{i=1}^{n}R_y(\phi_i)R_x(\theta_i)\right)
\end{equation}

    where $L$ denotes the number of quantum layers. We consider the depth of quantum layers to be 2, i.e., $L=2$.
    The final qubit state will be after QNN acts on the initial state is:
    \vspace{-8pt}
    $$\ket{\psi_{final}} = U(\theta)\ket{\psi_0}$$
    
    \item Finally, the expectation value of qubits are measured to get the output. For measurement we use PauliX, PauliY PauliZ and customized Hermitian measurement observables as discussed above.

\end{enumerate}




\vspace{-20pt}
\subsection{\textbf{Optimization Landscape Analysis}}

We conduct a detailed analysis of the optimization landscape of the QNNs under investigation by plotting the cost function relative to the network parameters, through which the optimizer navigates to find the optimal solution. Our analysis identifies and examines local and global minima within these landscapes.
Landscapes with abundant and broader global minima are generally conducive to effective optimization, facilitating the optimizer's path to the optimal solution. Conversely, landscapes dominated by numerous local minima or extensive flat regions are less favorable for optimization, posing greater challenges in reaching the optimal solution.

\vspace{-10pt}

\subsection{\textbf{Training Analysis}}
The QNNs are then trained for the problem defined in Eq. \ref{eq:CF}, under both noisy and ideal conditions. 
The training analysis systematically examines the convergence behavior of the cost function across a predetermined set of training iterations. This involves scrutinizing the evolution and stabilization of the cost function during training, providing insights into the effectiveness and efficiency of the learning process over time.

\vspace{-10pt}
\section{Experimental Setup} \label{sec:exp_setup}
\vspace{-5pt}
The QNNs constructed in previous section are trained to learn an Identity function. The cost function in this context is described Eq. \ref{eq:CF}: 

\vspace{-10pt}

\begin{equation} \label{eq:CF}
    C = \bra{\psi(\theta)} (I-\ket{00\ldots0}\bra{00\ldots0}) \ket{\psi(\theta)} = 1- p_{\ket{00\ldots0}} 
\end{equation}


We typically want to maximize the probability of all the qubits being in $\ket{0}$ state. 
%
The QNNs are trained for $50$ iterations. Adam optimizer with a learning rate of $0.1$ is used for the optimization. The experiments are performed using Pennylane \cite{Bergholm:2018}.   

\vspace{-10pt}


\section{Results and Discussion}
\vspace{-8pt}

\subsection{\textbf{Phase Damping}}
\vspace{-4pt}

We first present a comparative analysis of the training dynamics of QNNs under both ideal (noise-free) and noisy conditions, focusing on phase damping as the primary noise source. 
We analyze changes in the optimization landscape of $4$-qubit QNNs, which include $16$ single and $6$ two-qubit gates, for different noise probabilities compared to the ideal scenario. This analysis is limited to $4$-qubit QNNs as their landscape behaviors typically reflect in training results, making it logical to discuss either training results or optimization landscapes.

Subsequently, we provide a training analysis where QNNs of varying sizes ($4, 6, 8,$ and $10$ qubits) are trained for the problem defined in Eq. \ref{eq:CF}, using various measurement observables and subjected to different types and probabilities of quantum noise.

\vspace{-8pt}

\subsubsection{\textbf{Comparative Analysis of Optimization Landscapes: Noisy vs. Ideal Conditions:}} \label{sec:results_global}
\begin{figure*}[h]
    \hspace{-40pt}
    \includegraphics[scale=0.27]{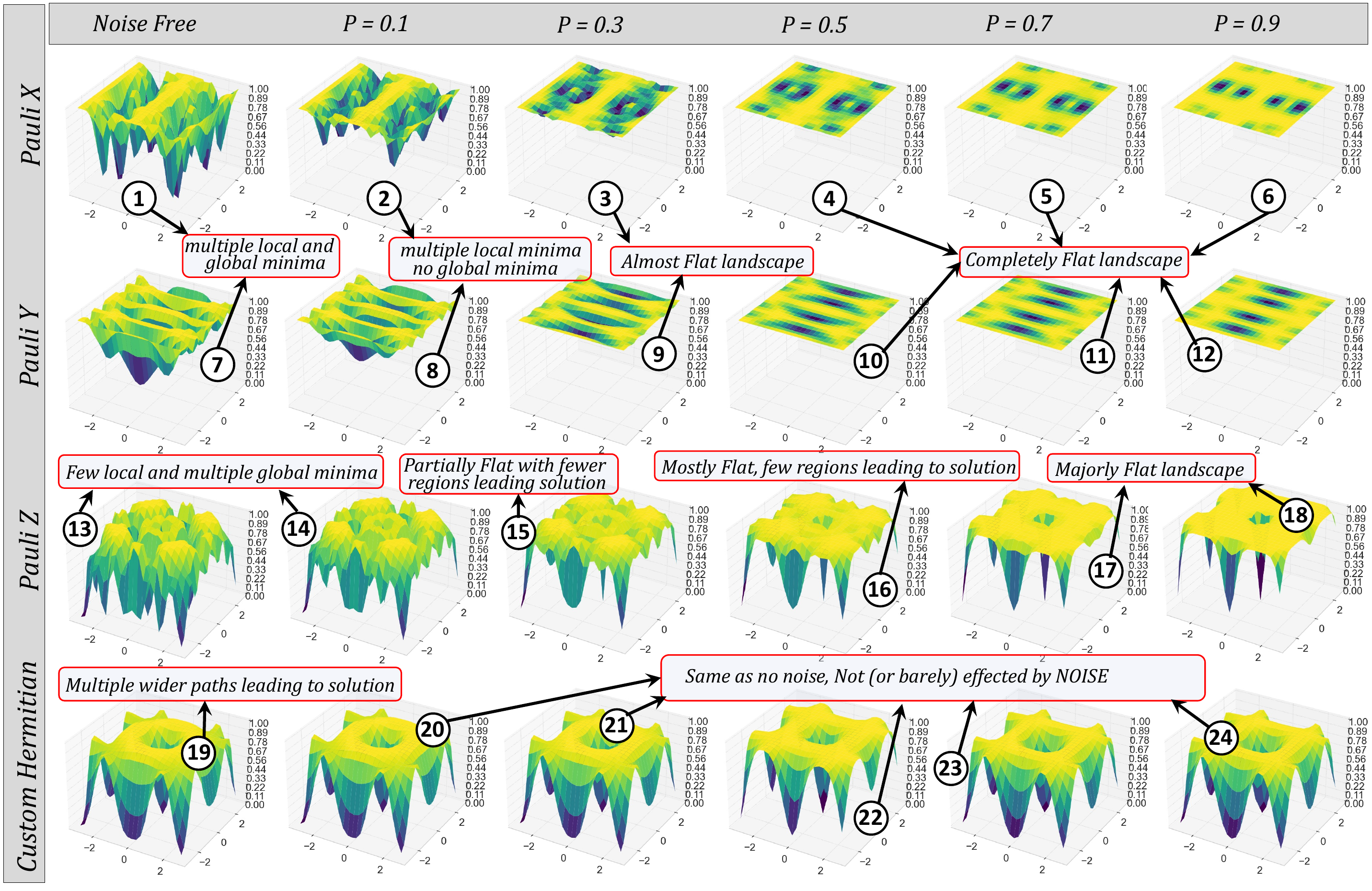}
    \vspace{-12pt}
    \caption{\footnotesize Optimization landscape of 4-qubit QNN in noise-free and under phase damping noisy environments. For almost all the observables, the landscapes in case of no noise have multiple wider regions containing the solution whereas the landscapes under noise tends to become flat as the noise probability increases except for customized Hermitian measurement observable which is not effected by noise. The x and y axis denotes parameters of quantum gates and z-axis denote the cost.}
    \label{fig:4Q_LS}
\end{figure*}

\vspace{-7pt}
The optimization landscapes for all qubit measurement observables, under both noisy (at varying noise probabilities) and ideal scenarios, are illustrated in Fig. \ref{fig:4Q_LS}. The feasibility of optimization varies across different observables. 
Below we analyze how the optimizations landscapes of QNNs with different measurement observables, alters under different quantum noise types. 

\vspace{-5pt}

\paragraph*{\textbf{Pauli Measurement Observables.}}
\vspace{-5pt}

In noise-free conditions, the optimization landscapes for 4-qubit QNNs using PauliX and PauliY observables, shown in labels \circled{1} and \circled{7} of Fig. \ref{fig:4Q_LS}, respectively, although contains \emph{multiple} local minima, but they also contain \emph{some}  global minima, indicating potential for effective optimization with proper hyperparameter tuning.
Introducing phase damping noise significantly changes these landscapes. At a low noise probability ($P=0.1$), the landscapes become flat, displaying multiple local minima but no global minima (labels \circled{2} and \circled{8} in Fig. \ref{fig:4Q_LS}). 
As noise probability increases from $P=0.3$ to $P=0.9$, the landscapes for both PauliX (labels \circled{3} to \circled{6}) and PauliY (labels \circled{9} to \circled{12}) observables become entirely flat, indicating early onset of BPs that severely limit QNN training efficiency even with fewer qubits, highlighting the significant impact of noise.

\hspace{-30pt}

For the PauliZ observable, the landscape's flattening with increasing noise is more gradual compared to PauliX and PauliY. In noise-free conditions, the PauliZ landscape is dynamic, with fewer local minima and multiple global minima (label \circled{13}). At $P=0.1$, the PauliZ landscape retains its dynamic structure with multiple solution pathways (label \circled{14}), maintaining training potential. At $P=0.3$, the landscape shows partial flattening, reducing optimization potential (label \circled{15}). As noise increases to $P=0.5$, the landscape flattens further, complicating the optimizer's task (label \circled{16}). At $P=0.7$ and $P=0.9$, the landscape becomes extensively flat, with few regions leading to global minima (labels \circled{17} and \circled{18}), significantly undermining training efficacy.

\vspace{-7pt}

\paragraph*{\textbf{Customized Hermitian Measurement Observable.}}
The customized Hermitian observable shows the most favorable optimization landscape under both noise-free and noisy conditions, surpassing PauliX, PauliY, and PauliZ observables. 
In noise-free scenarios, QNNs with this observable exhibit optimal landscapes with no local minima and expansive pathways to the solution (label \circled{19} in Fig.\ref{fig:4Q_LS}).
Introducing phase damping noise shows remarkable resilience in the customized Hermitian landscape, which remains unaffected regardless of the noise probabilities (labels \circled{20} to \circled{24} in Fig.\ref{fig:4Q_LS}). This resilience underscores its exceptional noise-resistant properties, affirming its significant potential for QNN training in noisy environments.

\subsubsection{\textbf{Comparative Analysis of Training in QNNs of Different Sizes:}}
\begin{figure*}[h]
    \hspace{-20pt}
    \includegraphics[scale=0.30]{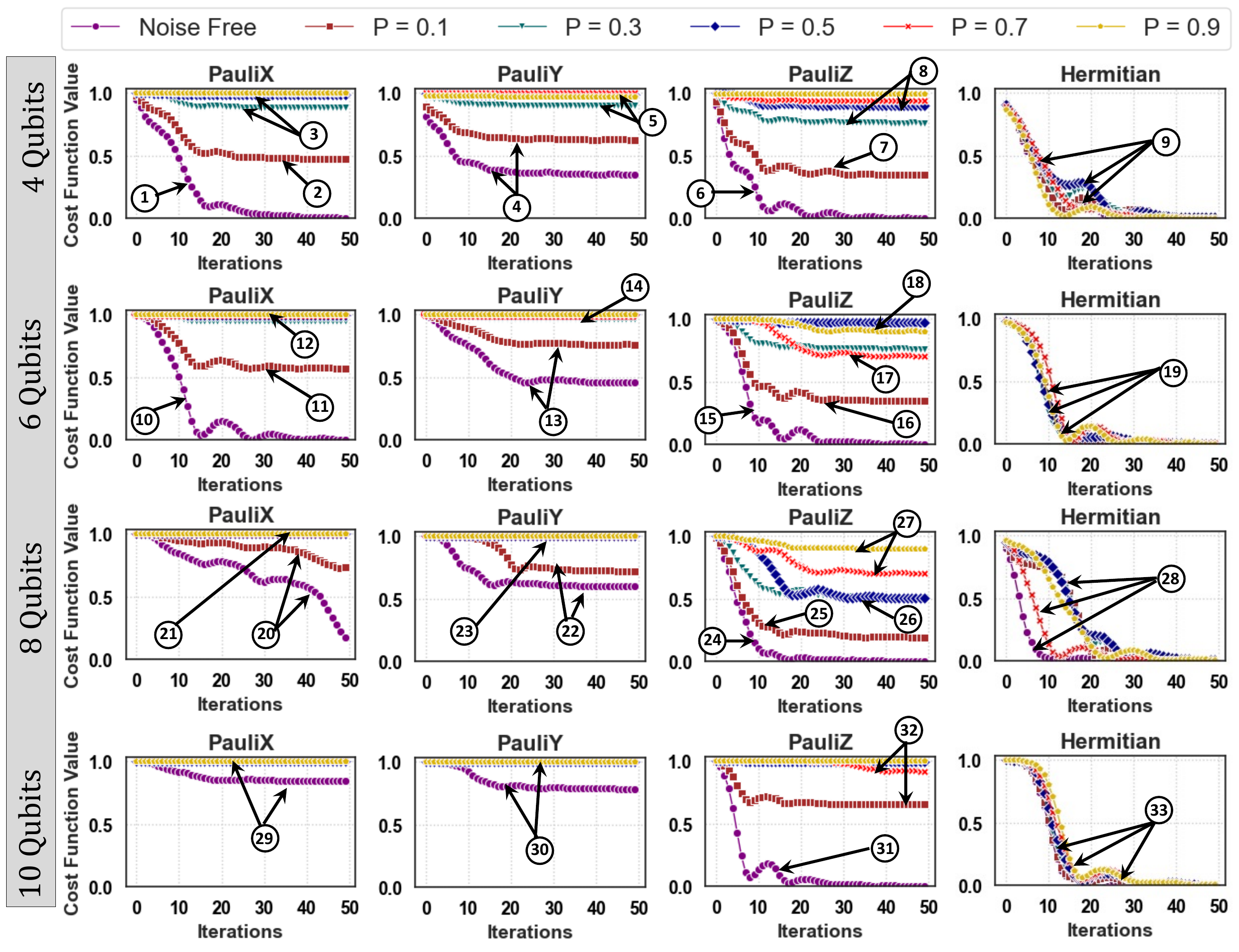}
    \vspace{-25pt}
    \caption{\footnotesize Training results of QNNs with different qubit measurement observables under \emph{Phase Damping} noise and noise-free environments. The most optimal training is achieved by customized Hermitian observable showing great resilience against noise.}
    \label{fig:training_PD}
\end{figure*}

\vspace{-10pt}
The training results for all QNNs with different measurement observables and different noise probabilities are presented in Fig. \ref{fig:training_PD}.
Below we separately discuss the training results of different sized QNNs:


\vspace{-2pt}

\paragraph{\textbf{4-Qubit QNNs:}}


\paragraph*{\textbf{Pauli Measurement Observables.}}
\vspace{-10pt}
In noise-free environments, QNNs using the PauliX measurement observable successfully converge to the solution (label \circled{1} in Fig. \ref{fig:training_PD}). However, introducing noise significantly degrades training performance. At a noise probability of $0.1$, QNNs exhibit suboptimal training (label \circled{2} in Fig. \ref{fig:training_PD}). As noise probabilities increase from $0.3$ to $0.9$, training performance further deteriorates, resulting in minimal to no training (labels \circled{3} to \circled{6} in Fig. \ref{fig:training_PD}). These results align with the optimization landscapes (labels \circled{1} to \circled{6} in Fig. \ref{fig:4Q_LS}).
%
%
%
%
QNNs using PauliY observables achieve suboptimal training even in noise-free conditions, similar to performance at a noise probability of $0.1$ (label \circled{4} in Fig. \ref{fig:training_PD}). With noise probabilities from $0.3$ to $0.9$, QNNs exhibit no training potential (label \circled{5} in Fig. \ref{fig:training_PD}). These results correspond with the optimization landscapes (labels \circled{7} to \circled{12} in Fig. \ref{fig:4Q_LS}).


In an ideal, noise-free environment, QNNs with PauliZ observables successfully converge (label \circled{6} in Fig. \ref{fig:training_PD}). At a noise probability of $0.1$, they show suboptimal training performance (label \circled{7} in Fig. \ref{fig:training_PD}). Higher noise probabilities result in negligible to no learning (label \circled{8} in Fig. \ref{fig:training_PD}). These results are consistent with the optimization landscapes (labels \circled{13} to \circled{18} in Fig. \ref{fig:4Q_LS}).


\vspace{-5pt}

\paragraph*{\textbf{Customized Hermitian Measurement Observable.}}

QNNs using the customized Hermitian observable demonstrate successful convergence in both noise-free and noisy environments. Unlike standard Pauli observables, this customized observable proves resilient to noise, maintaining effective training regardless of noise probabilities (label \circled{9} in Fig. \ref{fig:training_PD}). These results are consistent with the optimization landscapes (labels \circled{19} to \circled{24} in Fig. \ref{fig:4Q_LS}).


\vspace{-5pt}

\paragraph{\textbf{6-Qubit QNNs:}}
Based on the architecture details in Section \ref{sec:method_QNN_arch}, the $6$-qubit QNNs comprise $24$ single-qubit parameterized gates and $10$ two-qubit entangling gates.

\paragraph*{\textbf{Pauli Measurement Observables.}} 
\vspace{-10pt}
In noise-free scenarios, $6$-qubit QNNs using PauliX and PauliY measurement observables demonstrate successful training (labels \circled{10} and \circled{13} in Fig. \ref{fig:training_PD}). However, introducing noise degrades performance, leading to suboptimal to no training (labels \circled{11}, \circled{12}, and \circled{14} in Fig. \ref{fig:training_PD}).
QNNs with PauliZ observables show successful training in noise-free settings (label \circled{15} in Fig. \ref{fig:training_PD}). However, the introduction of noise causes a slight decline in training performance, resulting in suboptimal performance at low noise probability of $P=0.1$ (labels \circled{16} in Fig. \ref{fig:training_PD}) and negligible to no training at higher noise probabilities (labels \circled{17} and \circled{18} in Fig. \ref{fig:training_PD}).


\vspace{-10pt}
\paragraph*{\textbf{Customized Hermitian Measurement Observable.}}
QNNs using the custom-designed Hermitian measurement observable exhibit successful training in both noise-free and noisy environments (label \circled{19} in Fig. \ref{fig:training_PD}). This resilience to noise indicates that the custom-designed Hermitian observable significantly enhances the robustness of QNNs against phase damping noise. Such an observable makes the training process more adaptive and tolerant to noise, showing substantial performance improvement compared to standard Pauli measurement observables. This capability underscores the potential of customized observables in advancing the practical applicability of QNNs in real-world noisy scenarios.



\paragraph{\textbf{8-Qubit QNNs:}}
The $8$-qubit QNNs comprise $32$ single-qubit parameterized gates and $14$ two-qubit entangling gates, as discussed in Section \ref{sec:method_QNN_arch}.

\paragraph*{\textbf{Pauli Measurement Observables.}}
\vspace{-8pt}
As the QNN size increases, QNNs using PauliX and PauliY observables start experiencing BPs even in noise-free scenarios, demonstrating suboptimal performance similar to lower noise probabilities ($P=0.1$) (labels \circled{20} and \circled{22} in Fig. \ref{fig:training_PD}). Higher noise probabilities result in no training at all (labels \circled{21} and \circled{23} in Fig. \ref{fig:training_PD}).
%
QNNs with PauliZ observables show successful training in noise-free conditions (label \circled{24} in Fig. \ref{fig:training_PD}). However, the introduction of noise at higher qubit count leads to a noticeable decline in training performance. As noise probability increases, performance progressively deteriorates, with lower noise levels associated with better training outcomes compared to higher noise probabilities (labels \circled{25} to \circled{27} in Fig. \ref{fig:training_PD}).


\paragraph*{\textbf{Customized Hermitian Measurement Observable.}}
\vspace{-7pt}
The increased size of QNNs does not compromise the resilience of custom-designed Hermitian observables against phase damping noise. QNNs using these custom observables continue to show remarkable resistance to noise, maintaining superior performance in both noisy and noise-free environments, regardless of noise probability (label \circled{28} in Fig. \ref{fig:training_PD}).


\vspace{-5pt}

\paragraph{\textbf{10-Qubit QNNs:}}
The $10$-qubit QNNs comprise $40$ single-qubit parameterized gates and $18$ two-qubit entangling gates, as discussed in Section \ref{sec:method_QNN_arch}.

\vspace{-5pt}

\paragraph*{\textbf{Pauli Measurement Observables.}}
For QNNs using PauliX and PauliY observables, the increase in qubit count significantly diminishes training capabilities, even in noise-free environments. This limitation is primarily due to the BP phenomenon, which becomes more pronounced with higher qubit counts. In noise-free settings, the QNNs exhibit negligible to \emph{No} training, especially compared to noisy settings at all probabilities (labels \circled{29} and \circled{30} in Fig. \ref{fig:training_PD}).
%
%
QNNs using PauliZ observables show successful training in noise-free conditions (label \circled{31} in Fig. \ref{fig:training_PD}). 
However, there is a significant decline in performance when exposed to noise (label \circled{32} in Fig. \ref{fig:training_PD}) with QNNs exhibiting negligible training potential at lower noise probability of $0.1$ and no training potential at higher probabilities.


\vspace{-7pt}

\paragraph*{\textbf{Custom Hermitian Measurement Observable.}}

Custom-designed Hermitian observables maintain their resilience to noise even as the number of qubits increases, a scenario typically exacerbating BP and noise challenges. 
As shown in label \circled{33} in Fig. \ref{fig:training_PD}, larger QNNs using these custom observables exhibit successful training and exceptional noise resistance, performing equally well in both noisy and noise-free environments. This noise insensitivity suggests that such observables can effectively mitigate noise impacts, a critical advantage for the scalability and practical application of QNNs in real-world quantum machine learning tasks on NISQ devices, where noise is unavoidable.




\subsection{\textbf{Phase Flip}}
\begin{figure*}[h]
    \hspace{-10pt}
    \includegraphics[scale=0.285]{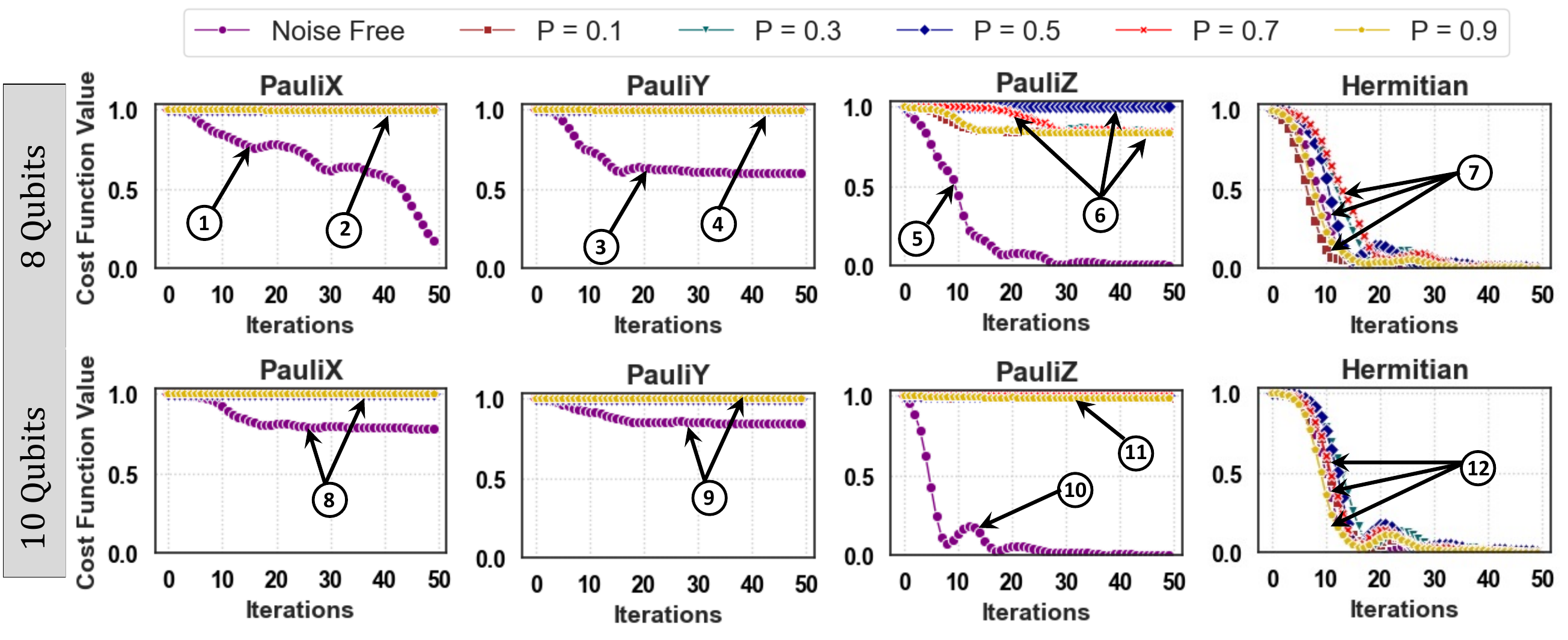}
    \vspace{-18pt}
    \caption{\footnotesize Training results of QNNs with different qubit measurement observables under \emph{Phase Flip} noise and noise-free environments. The most optimal training is achieved by customized Hermitian observable.}
    \label{fig:training_PF}
\end{figure*}

\vspace{-10pt}
We now present the training analysis of $8$ and $10$-qubit QNNs in both ideal and noisy settings, with phase flip as the primary source of noise. 
The QNNs are trained for the problem defined in Eq. \ref{eq:CF}, and the results are presented in Fig. \ref{fig:training_PF}. 
Here, we focus on larger QNNs ($8$ and $10$ qubits) because noise effects are more pronounced in more expressive networks. 
Demonstrating effective noise-resilient strategies in larger QNNs suggests potential benefits for smaller circuits.


\vspace{-8pt}
\paragraph{\textbf{8-Qubit QNNs:}}
\paragraph*{\textbf{Pauli Measurement Observables.}}
\vspace{-13pt}
In noise-free scenarios, QNNs using the PauliX observable show significant training potential (label \circled{1} in Fig. \ref{fig:training_PF}), which could improve further with hyperparameter optimization. 
QNNs using the PauliY observable train to some extent but quickly encounter BPs, resulting in suboptimal performance (label \circled{3} in Fig. \ref{fig:training_PF}). 
Introducing noise significantly declines training performance for both PauliX and PauliY observables (labels \circled{2} and \circled{4} in Fig. \ref{fig:training_PF}), indicating quick BP occurrence under noise influence.

\vspace{-15pt}

\vspace{-20pt}
QNNs using the PauliZ observable achieve significant training and successfully converge to the solution in noise-free settings (label \circled{5} in Fig. \ref{fig:training_PF}). However, noise introduction leads to suboptimal or no training (label \circled{6} in Fig. \ref{fig:training_PF}).



\paragraph*{\textbf{Customized Hermitian Measurement Observable.}} 
\vspace{-7pt}
The custom Hermitian observable demonstrates remarkable resilience against noise, sustaining effective training across all noise probability levels (label \circled{7} in Fig. \ref{fig:training_PF}). This highlights the potential benefits of employing custom-designed Hermitian observables in QNN architectures.

\vspace{-10pt}


\paragraph{\textbf{10-Qubit QNNs:}}
The performance of $10$-qubit QNNs is similar to that of $8$-qubit QNNs. However, a notable distinction is observed in the noise-free environment. While $8$-qubit QNNs with PauliX observables show considerable training potential, expanding to $10$ qubits exacerbates the BP problem, significantly reducing performance even without noise (label \circled{8} in Fig. \ref{fig:training_PF}). For other measurement observables, the performance remains similar to that of $8$-qubit QNNs across both ideal and noisy conditions (labels \circled{9} to \circled{12} in Fig. \ref{fig:training_PF}).


\vspace{-15pt}

\subsection{\textbf{Amplitude Damping}}
\begin{figure*}[h]
    \hspace{-20pt}
    \includegraphics[scale=0.29]{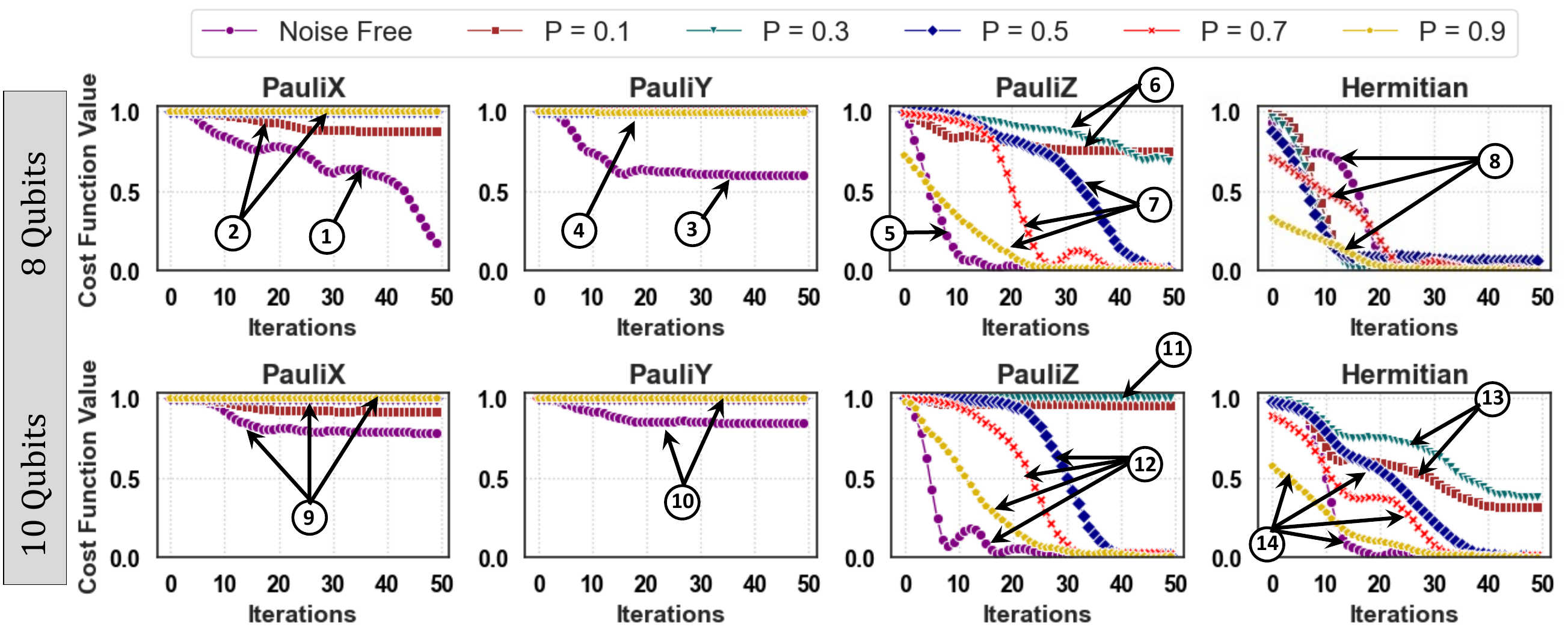}
    \vspace{-25pt}
    \caption{\footnotesize Training results of QNNs with different qubit measurement observables under \emph{Amplitude Damping} noise and noise-free environments. The most optimal training is achieved by customized Hermitian observable.}
    \label{fig:training_AD}
\end{figure*}

\vspace{-9pt}
We now present the training analysis of $8$ and $10$-qubit QNNs in both ideal and noisy settings, with \emph{amplitude damping} as the primary source of noise. The QNNs are trained for the problem defined in Eq. \ref{eq:CF}, with results shown in Fig. \ref{fig:training_AD}. An interesting observation here is that unlike phase damping and phase flip noise, where performance worsens with increasing noise probability, amplitude damping noise initially reduces the performance at low noise levels but improves at higher probabilities, approaching ideal conditions.

\vspace{-18pt}

%

\vspace{-30pt}
\paragraph{\textbf{8-Qubit QNNs:}}

\paragraph*{\textbf{Pauli Measurement Observables.}} 
\vspace{-12pt}
In noise-free settings, QNNs with PauliX observables show significant training potential (label \circled{1} in Fig. \ref{fig:training_AD}). 
Under noisy conditions, however, these QNNs exhibit negligible training potential regardless of the noise probability (label \circled{2} in Fig. \ref{fig:training_AD}). 
QNNs with PauliY observables show suboptimal training in ideal conditions (label \circled{3} in Fig. \ref{fig:training_AD}), and fail to train effectively in noisy conditions (label \circled{4} in Fig. \ref{fig:training_AD}).
%
%
QNNs with PauliZ observables successfully converge in noise-free settings (label \circled{5} in Fig. \ref{fig:training_AD}). In the presence of noise, performance deteriorates significantly at lower noise probabilities ($P=0.1$ and $0.3$) (label \circled{6} in Fig. \ref{fig:training_AD}). Notably, at higher noise probabilities ($P=0.5$ to $0.9$), training performance improves, and the model tends to converge towards the solution (label \circled{7} in Fig. \ref{fig:training_AD}).


\paragraph*{\textbf{Customized Hermitian Measurement Observable.}}
\vspace{-10pt}
The custom Hermitian observable shows great resilience to amplitude damping noise. QNNs with all the noise probabilities successfully converge to the solution (label \circled{8} in Fig. \ref{fig:training_AD}).


\paragraph{\textbf{10-Qubit QNNs:}}

\vspace{-8pt}
\paragraph*{\textbf{Pauli Measurement Observables.}}
\vspace{-11pt}
The $10$-qubit QNNs with PauliX and PauliY observables perform significantly worse than 8-qubit QNNs, even in noise-free environments, due to the emergence of the BP problem as qubit count increases (labels \circled{9} and \circled{10} in Fig. \ref{fig:training_AD}).
%
At lower noise probabilities, $10$-qubit QNNs with PauliZ observables show no training potential. As noise probability increases, training performance improves (labels \circled{11} and \circled{12} in Fig. \ref{fig:training_AD}).

\vspace{-7pt}
\paragraph*{\textbf{Customized Hermitian Measurement Observable.}}
The $10$-qubit QNNs with custom Hermitian observables experience a slight performance decline at lower noise probabilities compared to $8$-qubit QNNs, likely due to the BP problem (label \circled{13} in Fig. \ref{fig:training_AD}). Nonetheless, these networks exhibit substantially better training potential than those with other observables. The resilience of custom Hermitian observables against amplitude damping noise at higher probabilities is significant (label \circled{14} in Fig. \ref{fig:training_AD}).



\vspace{-10pt}

\section{Conclusion}
\vspace{-10pt}
This paper delves into the complexities of training Quantum Neural Networks (QNNs) in the presence of quantum noise, a crucial aspect for the advancement of Noisy Intermediate-Scale Quantum (NISQ) devices. We demonstrated that Barren Plateaus (BPs), defined by exponentially vanishing gradients, occur earlier (even at smaller qubit count) in noisy environments compared to ideal, noiseless conditions. This early onset of BPs poses a significant challenge to the training efficacy of QNNs, necessitating innovative approaches to mitigate these effects.
Our investigation focuses on the impact of different qubit measurement observables, i.e., PauliX, PauliY, PauliZ, and a custom-designed Hermitian observable, on QNN's performance under various noise conditions, including Phase Damping, Phase Flip, and Amplitude Damping. 
Our findings reveal a marked vulnerability of QNNs utilizing standard Pauli observables to the early emergence of BPs, even with a minimal qubit count of $4$ qubits in noisy environments. 
In contrast, the custom Hermitian measurement observable showed exceptional resilience, maintaining consistent trainability up to $10$ qubits across all types of quantum noise.

Our results underscore the pivotal role of qubit measurement observable selection in enhancing QNN robustness and performance against BPs under noisy conditions. By highlighting the superior resilience of the custom Hermitian observable, this study provides a strategic pathway for designing more noise-resilient QNNs, and offers significant potential for improving QNN training efficacy and contributes to the broader field of quantum machine learning by addressing fundamental challenges posed by quantum noise. 
We aim to extend our research to explore and refine such strategies for a diverse range of quantum noise, paving the way for more robust and effective quantum computing applications.

\vspace{-15pt}
\section*{Acknowledgements}
\vspace{-10pt}
This work was supported in part by the NYUAD Center for Quantum and
Topological Systems (CQTS), funded by Tamkeen under the NYUAD Research
Institute grant CG008.

\vspace{-22pt}

\end{spacing}

\begin{spacing}{0.2}
\small
\bibliographystyle{splncs04}
\bibliography{main} 
\end{spacing}




\end{document}